\documentclass[twocolumn,showpacs,preprintnumbers,amsmath,amssymb,aps,prl,superscriptaddress]{revtex4-2}
\usepackage{epsfig}
\usepackage{graphicx}% Include figure files
\usepackage{dcolumn}% Align table columns on the decimal point
\usepackage{bm}% bold math
\usepackage{color} % \color{red} \color{black}
\usepackage{ulem} % \sout - single out

\usepackage{xspace}

\newcommand{\EF}{$E_{\rm F}$\xspace}
\newcommand{\bSmS}{$b$-SmS\xspace}
\newcommand{\gSmS}{$g$-SmS\xspace}
\newcommand{\dSm}{Sm$^{2+}$\xspace}
\newcommand{\tSm}{Sm$^{3+}$\xspace}
\newcommand{\cf}{$c$-$f$\xspace}
\newcommand{\OC}{$\sigma_1(\omega)$\xspace}
\newcommand{\R}{$R(\omega)$\xspace}
\newcommand{\IV}{$I$-$V$\xspace}

\newcommand{\hw}{$\hbar\omega$\xspace}

\newcommand{\Neff}{$N_{eff}$\xspace}

\begin{document}

\title{
Current-Induced Metallization and Valence Transition in Black SmS%: \\Optical and Photoelectrical Studies
}
\author{Shin-ichi Kimura}
\email{kimura.shin-ichi.fbs@osaka-u.ac.jp}
\affiliation{Graduate School of Frontier Biosciences, Osaka University, Suita, Osaka 565-0871, Japan}
\affiliation{Department of Physics, Graduate School of Science, Osaka University, Toyonaka, Osaka 560-0043, Japan}
\affiliation{Institute for Molecular Science, Okazaki, Aichi 444-8585, Japan}
\author{Hiroshi Watanabe}
\affiliation{Graduate School of Frontier Biosciences, Osaka University, Suita, Osaka 565-0871, Japan}
\affiliation{Department of Physics, Graduate School of Science, Osaka University, Toyonaka, Osaka 560-0043, Japan}
\author{Shingo Tatsukawa}
\affiliation{Department of Physics, Graduate School of Science, Osaka University, Toyonaka, Osaka 560-0043, Japan}
\author{Takuto Nakamura}
\affiliation{Graduate School of Frontier Biosciences, Osaka University, Suita, Osaka 565-0871, Japan}
\affiliation{Department of Physics, Graduate School of Science, Osaka University, Toyonaka, Osaka 560-0043, Japan}
\author{Keiichiro Imura}
\thanks{Present address: Institute of Liberal Arts and Sciences, Nagoya University, Nagoya, Aichi 464-8601, Japan}
\affiliation{Department of Physics, Graduate School of Science, Nagoya University, Nagoya Aichi 464-8602, Japan}
\author{Hiroyuki S. Suzuki}
\affiliation{Institute for Solid State Physics, The University of Tokyo, Kashiwa, Chiba 277-8581, Japan}
\author{Noriaki K. Sato}
\thanks{Present address: Center for General Education, Aichi Institute of Technology, Toyota, Aichi 470-0392, Japan}
\affiliation{Department of Physics, Graduate School of Science, Nagoya University, Nagoya Aichi 464-8602, Japan}
\date{\today}
\begin{abstract}
A strongly-correlated insulator, samarium mono-sulfide (SmS), presents not only the pressure-induced insulator-to-metal transition (IMT) with the color change from black to golden-yellow but also current-induced IMT (CIMT) with negative resistance.
To clarify the origin of the CIMT of SmS, the electronic structure change has been investigated by optical reflectivity and angle-integrated photoelectron spectra by applying an electric current.
At lower temperatures than about 100~K, where the nonlinear $V$-$I$ curve has been observed, the carrier density rapidly increases, accompanied by decreasing relaxation time of carriers with increasing current.
Then, the direct gap size increases, and the mean valence changes from Sm$^{2+}$-dominant SmS to the mixed-valent one with increasing current.
These results suggest that the CIMT originates from increasing the Sm~$4f$-$5d$ hybridization intensity induced by the applied current.
\end{abstract}

%
%\pacs{71.27.+a, 78.20.-e}% PACS, the Physics and Astronomy
%%%%%%%%%%%%%%%%%%%%%%%%%%%%%%%%%%%%%%%%%%%%%%%%%%%%%%%%%%%%
\maketitle
%
%%%%%%%%%%%%%%%%%%%%%%%%%%%%%%%%%%%%%%%%%%%%%%%%%%%%%%%%%%%%
%\section{Introduction}
%
Insulator-to-metal transition (IMT) of solids has been one of the critical topics of interest for a long time because of the drastic change in electrical resistivity that is useful for microscopic switching~\cite{Mott1968, Mott2004}.
IMT has been observed in many transition-metal compounds and organic conductors, in which they accompany a Jahn-Teller effect~\cite{Han2000}, a charge/orbital ordering~\cite{Moritomo1997}, and a lattice distortion due to the creation of charge/spin-density waves~\cite{Gruner1994}.
On the other hand, in rare-earth compounds, for instance, in SmB$_6$ and YbB$_{12}$ and others, metallic property at higher temperatures changes to a Kondo insulator at low temperatures with a tiny energy gap due to the hybridization between conduction and $f$ states (\cf hybridization) by the Kondo effect~\cite{Kasuya1994}.
The nonmagnetic insulating property of EuO at high temperatures changes to a ferromagnetic metal due to the indirect exchange interaction~\cite{Lee1984, Miyazaki2009}.
These rare-earth compounds do not show a lattice distortion because of the weak crystal electric field of $4f$ electrons.
Samarium mono-sulfide (SmS), the title compound, also shows IMT by applying pressure without lattice distortion as other rare-earth compounds.

SmS with NaCl-type crystal structure has the pressure-induced first-order phase transition from the black-colored semiconductor (\bSmS) to the golden-yellow-colored semimetal (\gSmS) (black-to-golden phase transition: BGT)~\cite{Jayaraman1970, Batlogg1974, Batlogg1976, Travaglini1984a, Matsubayashi2007b, Matsubayashi2007a}.
BGT can also be produced by applying chemical pressure of the substituting trivalent yttrium (Y) ions with a smaller ionic radius to \tSm ions~\cite{Penney1975, Guntherodt1976, Kaneko2014, Yokoyama2019a}.
The BGT accompanies the valence transition from \dSm to \tSm, and the lattice constant decreases by about 5~\% due to the shrinking of the ionic radius without change in the crystal structure~\cite{Jarrige2013, Imura2020}.
At BGT, the electrons' contribution of Sm ions changes from \dSm~$4f^6$ in \bSmS to \tSm~$4f^5+c$ ($c$: conduction electron) in \gSmS by the BGT, so the appearance of the carriers is considered the origin of the IMT.
Since the valence change and the shrinking of the lattice constant coincidently occur, however, more than 50 years after its discovery, the origin of BGT of this material is still under debate~\cite{Watanabe2021}.
One of the proposed ideas for the origin of BGT is the BEC--BCS transition of Sm~$4f$-$5d$ exciton~\cite{Varma1976}.
The origin of the BEC--BCS transition is the spontaneous emergence of electron-hole pairs, namely exciton, due to the decrease of the energy gap by applying pressure.
If so, creating the $4f$-$5d$ exciton due to other methods is expected to make BGT.

Recently, Ando {\it et al.} reported that the voltage ($V$) as a function of current ($I$) changes non-linearly from the high $dV/dI$ at low $I$ to the low $dV/dI$ at high $I$ via a negative $dV/dI$ at lower temperatures than 100~K, namely a current-induced IMT (CIMT). 
It is discussed to be related to the BEC--BCS transition of the $4f$-$5d$ excitons~\cite{Ando2020}.
The change of the $dV/dI$ value reflects the difference in the parameters of the conduction electrons, the carrier density ($N$), and/or the mobility ($\mu$), which is related to the electronic structure change.
However, the origin of the CIMT has yet to be revealed.

In this Letter, to investigate the origin of the CIMT of \bSmS from the view of the electronic structure, 
we measured reflectivity (\R) spectra in the broad energy region from the terahertz (THz) to the vacuum-ultraviolet and angle-integrated photoelectron (PE) spectra by applying electric currents.
As a result, the carrier density rapidly increases with decreasing mobility by applying current in the negative resistance region.
In addition, the energy gap size increases, and the mean valence of Sm-ions increases with increasing current.
However, the carrier density and the mean valence after CIMT do not reach those of \gSmS, suggesting a different mechanism of BGT or a precursor of BGT.
The obtained results suggest that the localized Sm~$4f$ electrons change to an itinerant character by applying current, similar to melting ice by flowing water, and the itinerant $4f$ electrons hybridize with the Sm~$5d$ conduction band.
The observed CIMT of \bSmS is concluded as a novel type of valence transition.

%%%%%%%%%%%%%%%%%%%%%%%%%%%%%%%%%%%%%%%%%%%%%%%%%%
%\section{Experimental}
%
Single crystalline samples of SmS were grown using the vertical Bridgman method with a high-frequency induction furnace: 
Pre-reacted materials starting 99.99~\% pure (4N) samarium and 6N powdered sulfur were sealed in a vacuum W crucible~\cite{Matsubayashi2007a}.
Cleaved samples with the size of about $1\times1\times1$~mm$^3$ were mounted on the sample folder with the current flow and cooled down to about 20~K for the \R measurement and to 30~K for the PE measurement.
The direct current up to 1.6~A was applied while maintaining a temperature range of up to 5~K to avoid temperature rise due to Joule heating and to prevent sample degradation.
Near-normal-incident \R spectra along the $(001)$ plane were acquired in a wide photon-energy range of 6~meV--10~eV to obtain optical conductivity (\OC)  spectra via the Kramers-Kronig analysis.
Infrared (IR) and THz measurements at the photon energy \hw regions of 50~meV--1.5~eV and 6--50~meV have been performed using IR and THz microscopes with synchrotron radiation at the beamline 6B of UVSOR-III, Institute for Molecular Science, Japan~\cite{Kimura2006,Kimura2007a}.
In the photon energy range of 2--10~eV, the \R spectra were measured using the synchrotron radiation setup at the beamline 3B of UVSOR-III~\cite{Fukui2014}.
PE spectra have been accumulated using a He-II discharge lamp with a monochromator (VG-Scienta VUV 5000+5040), a hemispherical analyzer (MBScientific A-1), and an own-developed cold sample finger with a current-flow system.
The base pressure for the PE measurement was less than $1\times10^{-8}$~Pa.

%%%%%%%%%%%%%%%%%%%%%%%%%%%%%%%%%%%%%%%%%%%%%%%%%

%\section{Results}

%%%%%%%%%%%%%% FIG. 1. IR spectra %%%%%%%%%%%%%%%%%%%%
\begin{figure*}[t]
\begin{center}
\includegraphics[width=0.8\textwidth]{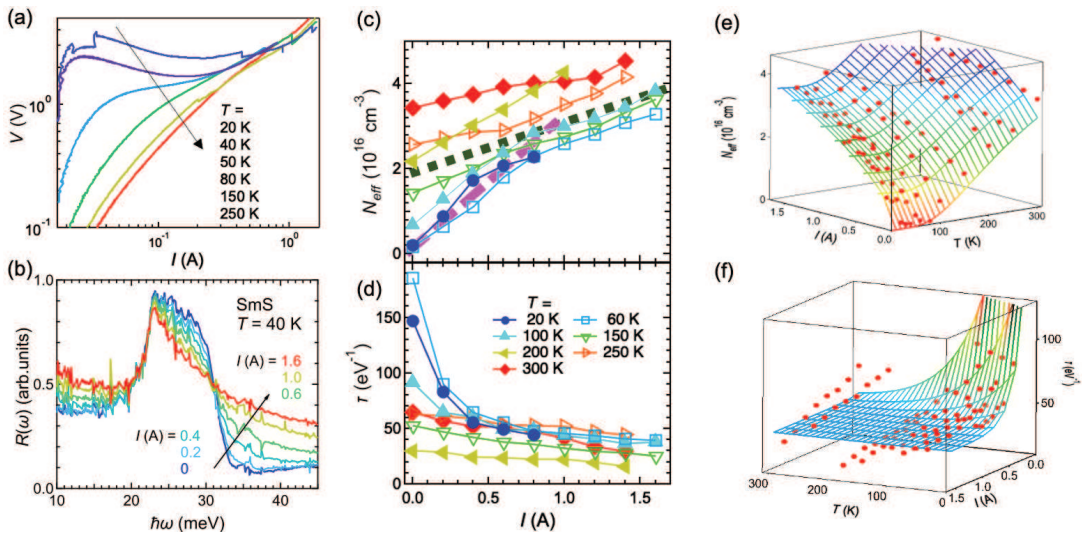}
\end{center}
\caption{
%(Color online)
(a) \IV curves of \bSmS at temperatures from 20 to 250~K.
(b) Current-dependent reflectivity (\R) spectra of \bSmS in the THz region at $T=40$~K.
(c, d) Temperature and external-current dependence of (c) the effective carrier density (\Neff) and (d) the relaxation time ($\tau$) of \bSmS 
evaluated by the Drude-Lorentz (DL) fitting of the THz \R spectra shown in (b) and at other temperatures.
(e, f) Same as (c) and (d), but all of the DL fitting data and schematic curved surfaces are also plotted as guides for the eye.
}
\label{fig:IR}
\end{figure*}
%%%%%%%%%%%%%%%%%%%%%%%%%%%%%%%%%%%%%%%

%\subsection{Temperature-dependent \IV curve}

Firstly, the temperature dependence of the \IV behavior has been checked.
Figure~\ref{fig:IR}a shows the \IV characteristics of \bSmS measured with increasing electric current at temperatures from 20 to 250~K.
The results are qualitatively consistent with that reported previously~\cite{Ando2020}.
The \IV curve changes to a monotonic increase at temperatures above 80~K.
Below 50~K, the \IV curves show a sizable non-linear behavior, and finally, below 40~K, negative $dV/dI$ appears in the current from 30 to 120~mA.
At 40~K, the $dV/dI$ value of $\sim10^2$~V/A at $I=20$~mA changes to $\sim2.5$~V/A at 1~A, {\it i.e.}, 50 times decreases, but at 80~K, it decreases about two times smaller.
This fact suggests that the conduction mechanism and the electronic structure change between the low-current and high-current regions at temperatures below 50~K.
On the other hand, at higher temperatures than 80~K, the electronic structure is unchanged.
The result at higher than 80~K is consistent with the recent electric-field-induced IMT results measured at higher than 77~K, which can be explained by Joule heating effect~\cite{Kishida2022}.

%%%%%%%%%%%%%
%\subsection{Current-induced carriers}

To obtain information on carriers at different currents, we derived the temperature and current dependences of the carrier density and the relaxation time from the Drude-Lorentz (DL) fitting of the \R spectra in the THz region~\cite{Mizuno2008,SM}.
As an example of the current-dependent THz \R spectra, spectra at $T=40$~K are shown in Fig.~\ref{fig:IR}b.
At $I = 0$~A, the spectrum is a typical insulating one, with a phonon peak at $\sim$25~meV, and the background intensity is low, suggesting a low carrier density.
With increasing current, the background intensity increases, accompanied by a rapid change in the phonon shape originating from the interface between the phonon and background absorptions.
After analyzing the spectral change using the DL fitting, we obtained the strong temperature- and current-dependent effective carrier density (\Neff) and the relaxation time ($\tau$) as shown in Figs.~\ref{fig:IR}c and \ref{fig:IR}d.
To show the temperature and current dependences more clearly, the three-dimensional plots are also shown in Figs.~\ref{fig:IR}e and \ref{fig:IR}f.
On the other hand, obtained parameters for the phonon peak showed little change with applied current as shown in Fig.~S2 in the Supplementary Material~\cite{SM}, suggesting the Joule heating effect is not large, consistent with the temperature kept within 5~K.
Therefore, the current dependence of \Neff and $\tau$ originates from the applied current itself.

As shown in Fig.~\ref{fig:IR}c, \Neff increases with increasing current at all temperatures.
At higher temperatures than 150~K, the rate of increase ($dN_{eff}/dI$) is almost constant at different temperatures ($\sim1.2\times10^{16}~{\rm cm}^{-3}{\rm A}^{-1}$, dotted lines in Fig.~\ref{fig:IR}c), but the intercept at 0~A increases with increasing temperature.
The \Neff at 0~A originates from the thermally excited carriers because the smallest (indirect) energy gap size of the material is about 0.1~eV, comparable to the thermal excitation energy of $k_{\rm B}T$.
The similar $dN_{eff}/dI$ value is also observed at higher currents than 0.6~A of lower temperatures than 100~K, suggesting the electronic structure in the region is consistent with the high-temperature range.

On the other hand, at lower currents than 0.6~A and lower temperatures than 100~K, the $dN_{eff}/dI$ value is about $3.2\times10^{16}$~cm$^{-3}$A$^{-1}$ (the dashed line in Fig.~\ref{fig:IR}c), which is about three times larger than that at the high-current region.
$\tau$ also enhances in the same temperature and current ranges.
Usually, $\tau$, as well as the electron mobility $\mu$, have a carrier-density dependence due to the electron-electron scattering, but, in the low-carrier-density region at around $10^{16}$~cm$^{-3}$, they do not usually change so much~\cite{Masetti1983,Reggiani2000}.
Therefore, the change of $\tau$ at low temperature and low current region suggests that the character of carriers and the electronic structure are modulated from the region of the low temperature and low current to that of the high temperature and high current.

%\subsection{Reflectivity and optical conductivity spectra in the broad energy region}

%%%%%%%%%%%%%% FIG. 2. OC spectra %%%%%%%%%%%%%%%%%%%%
\begin{figure}[t]
\begin{center}
\includegraphics[width=0.4\textwidth]{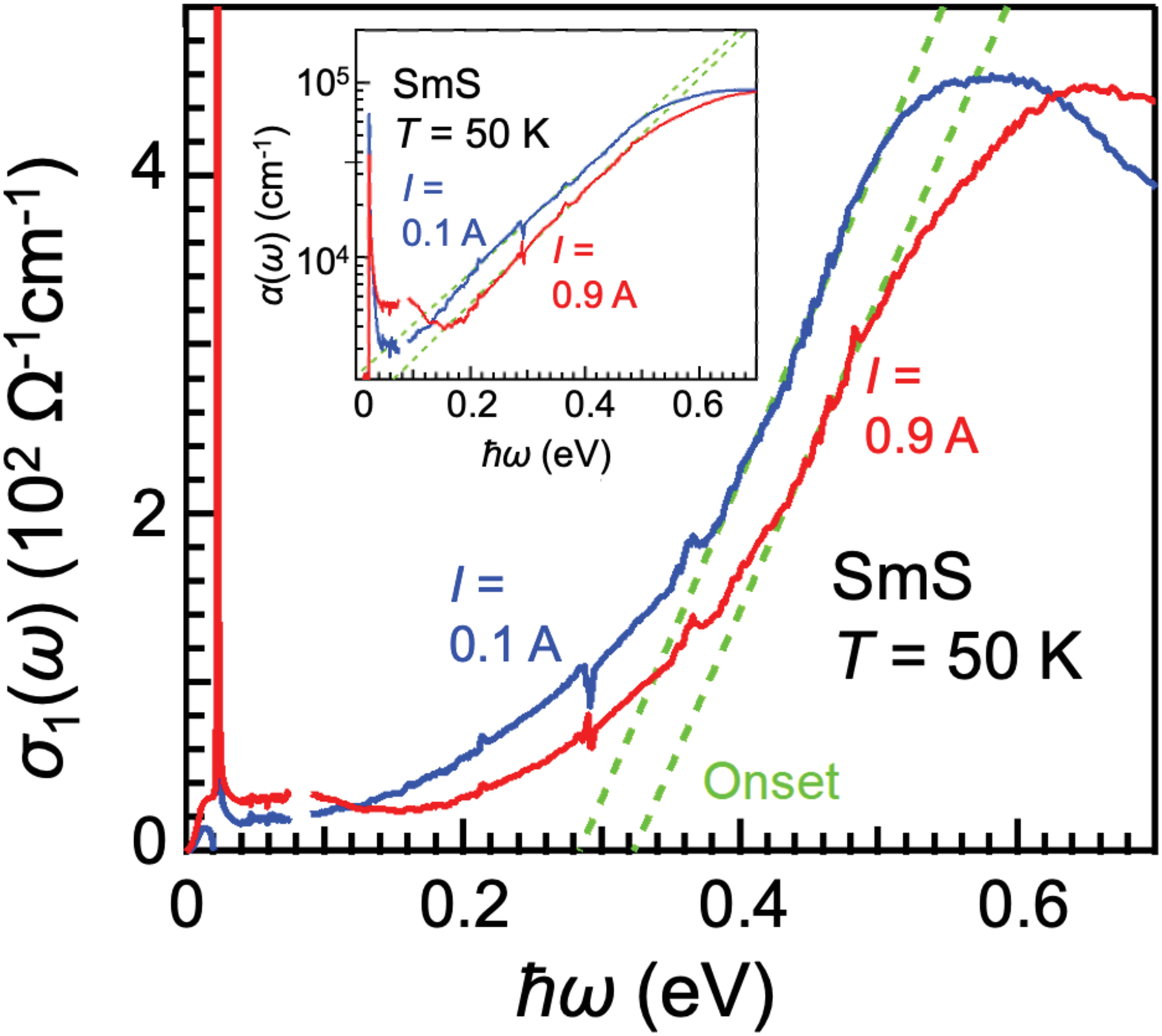}
\end{center}
\caption{
%(Color online)
Current-dependent optical conductivity(\OC) spectra of \bSmS in \hw below 0.7~eV at $T=50$~K, derived from the Kramers-Kronig analysis (KKA) of the reflectivity (\R) spectra.
In the low current of $I=0.1$~A, the onset of the reading edge is about 0.28~eV, whereas it shifts to about 0.32~eV in the high current of 0.9~A.
The inset shows that the absorption edges in the absorption ($\alpha(\omega)$) spectra obtained from the KKA obey the $\exp(\hbar\omega)$-law, suggesting evidence of a direct gap from the Urbuch rule of exciton peaks~\cite{Urbach1953}.
A sharp peak at about 20~meV originates from a TO phonon that is the same as the peak shown in Fig.~\ref{fig:IR}b.
}
\label{fig:OC}
\end{figure}
%%%%%%%%%%%%%%%%%%%%%%%%%%%%%%%%%%%%%%%

To clarify the electronic structure change by the current application, the current-dependent \OC spectra at $T = 50$~K are shown in Fig.~\ref{fig:OC}.
At the low-current region of $I=0.1$~A, a peak appears at about 0.56~eV originating from the exciton absorption at the direct gap at the $X$ point~\cite{Mizuno2008}, as evidenced in the inset of Fig.~\ref{fig:OC}.
By applying a current of $\sim$0.9~A, the peak at $\sim0.64$~eV shifts to the high-energy side.
The onset of the reading edge also increases by about 40~meV from about 0.28~eV to 0.32~eV with increasing current.
These results indicate the enlargement of the energy gap size.
The obtained \OC spectrum at high current is completely different from that of \gSmS, in which there is a Drude-like spectral shape with a plasma edge at about 3~eV~\cite{Kirk1972, Batlogg1976, Travaglini1984a}.
Therefore, the electronic structure of the high-current phase is not the same as that of \gSmS.
Instead, the high-current spectrum in the low-energy region below 0.2~eV is similar to the spectrum at the pressure of 0.6~GPa, just below the boundary of BGT~\cite{Mizuno2008}.
However, the exciton peak shifts to the low-energy side by a pressure application, which trend is the opposite behavior of the current application.
The pressure-induced energy-gap narrowing below 0.6~GPa indicates that the material is still in the black phase at the pressure; therefore, the thermally excited carrier appears.
However, in the high-current spectrum, the energy gap size increases, even though the carrier density increases with increasing current.
This fact suggests that the carriers in the high-current region do not originate from thermal excitation.
One scenario for the energy-gap enlargement at high currents is the appearance of the bonding and anti-bonding states of the hybridization between the Sm~$4f$ valence band and the $5d$ conduction band.
In the low-current region, a direct gap is expected at the $X$ point in the Brillouin zone~\cite{Kimura2008b}.
Then, since the Sm~$4f$ states are isolated, the Sm~$5d$ conduction band, which orbital the carriers go through, is not hybridized.
At the high-current region, the isolated $4f$ electrons change to itinerant; the $4f$ electrons can be hybridized with the carriers, {\it i.e.}, the hybridization bands, namely bonding and anti-bonding states of the $4f$ states and $5d$ conduction bands, are created.
The energy gap between the bonding and anti-bonding states is enlarged with increasing the hybridization intensity.
In addition, the mean valence of Sm ions is expected to be changed from $2+$ to $2+\delta$ due to the $4f$-$5d$ hybridization.
Then, the carrier density released from the Sm~$4f$ states can increase with increasing the hybridization intensity.

%%%%%%%%%%%%%%%%%%%%%%%%%%%%%%%%%%%%%%%

%\subsection{\red{Current-induced valence change}}

%%%%%%%%%%%%%% FIG. 3. PES %%%%%%%%%%%%%%%%%%%%
\begin{figure}[t]
\begin{center}
\includegraphics[width=0.4\textwidth]{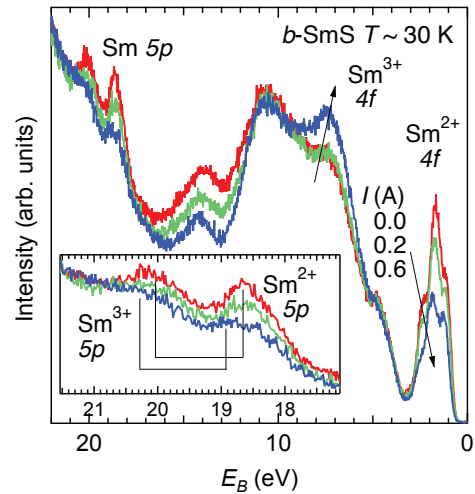}
\end{center}
\caption{
%(Color online)
Current-dependent angle-integrated photoelectron (PE) spectra in the valence band energy region as a function of binding energy ($E_B$).
The PE spectra are accumulated at about 30~K using the He-II line ($h\nu=40.8$~eV) as a light source.
(Inset) Enlarge figure of the Sm~$5p$ core level.
%\colorbox{yellow}{This figure must be updated.}
}
\label{fig:PES}
\end{figure}
%%%%%%%%%%%%%%%%%%%%%%%%%%%%%%%%%%%%%%%

To check the hybridization between the Sm~$4f$ and $5d$ states at high currents, we measured the current-dependent valence band PE spectra as shown in Fig.~\ref{fig:PES}.
The figure shows that the PE spectrum has strong current dependence, {\it i.e.}, the intensity of the Sm$^{2+}~4f$ multiplet peaks in the binding energy range below 3~eV decreases, and in contrast, the Sm$^{3+}$ peak at about 8~eV increases with increasing current.
Similarly, as shown in the inset, the intensity of the Sm$^{2+}~5p$ peaks decreases, and that of the Sm$^{3+}~5p$ peaks increases with the current application.
These results strongly support that the applied current increases the Sm mean valence from $2+$ to $2+\delta$.
Therefore, the scenario of the appearance of the hybridization state between Sm~$4f$ and $5d$ at the high-current region is plausible.

%%%%%%%%%%%%%% FIG. 4. Expected electronic structure change %%%%%%%%%%%%%%%%%%%%
\begin{figure}[t]
\begin{center}
\includegraphics[width=0.45\textwidth]{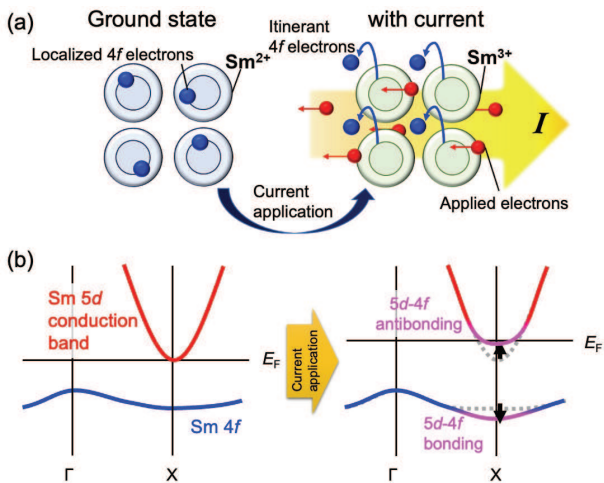}
\end{center}
\caption{
%(Color online)
The scenario of the current-induced metallization and valence transition of \bSmS.
(a) Real space description of the current-induced metallization. 
Localized $4f$ electrons are induced to be itinerant by applied currents, like melting ice by flowing water.
(b) Reciprocal lattice description of the current-induced metallization and valence transition.
After changing to the itinerant $4f$ electrons at high currents, they hybridize to carriers, namely $4f$-$5d$ hybridization.
Then the bottom of the conduction band becomes flat. 
The relaxation time $\tau$, which corresponds to the mobility ($\mu=e\tau/m$, where $e$ is the electron charge and $m$ the electron mass), becomes shorter than that in the low-current case.
After the appearance of the hybridization, the Sm valence changes from $2+$ to $2+\delta$.
Then the Fermi level (\EF) moves to the higher energy side.
}
\label{fig:ElectronicStructure}
\end{figure}
%%%%%%%%%%%%%%%%%%%%%%%%%%%%%%%%%%%%%%%

To explain all of the experimental evidence, the lowering $\tau$, the energy gap enlargement and the valence transition from $2+$ to $2+\delta$ by current application, we propose a model as shown in Fig.~\ref{fig:ElectronicStructure}.
At the ground state without current application, the Sm$^{2+}~4f$ states are localized, and the energy gap between the Sm~$4f$ and $5d$ states opens~\cite{Kimura2008b}.
Since the $5d$ conduction band is a free-electron-like band and the Sm ions are localized and non-magnetic divalent, 
the carriers' relaxation time $\tau$ is long because of the lack of strong interactions.
With applying current, the flowing electrons push the localized $4f$ electrons out from their localized sites by an electron-electron interaction because the energy level of the localized $4f$ electrons is very close to \EF, and then the $4f$ electrons become itinerant.
The itinerant $4f$ electrons can hybridize with carriers in the Sm~$5d$ states, so the $4f$-$5d$ hybridization state appears.
Carriers in the hybridization band have a shorter relaxation time because of the strong interaction.

Similar phenomena to the CIMT and valence transition on \bSmS have not been observed, to the best of our knowledge, even though current-induced phenomena, such as a N\'eel vector switching~\cite{Wu2022}, phase transitions in cuprate~\cite{Naito2021}, and a magnetic thin film~\cite{Kim2008}, 
has recently been reported.
Also, electric-field-induced metallization on Ca$_2$RuO$_4$~\cite{Nakamura2013,Zhang2019c} is similar to our observation, but the electric field in our study is not so high.
Our observation of the current-induced metallization and valence transition completely differs from such works.
This phenomenon may be related to other anomalous phenomena, such as pressure-induced BGT, alkali-metal-induced surface valence transition~\cite{Nakamura2022a}, photo-induced band shift~\cite{Chen2022}, and the BEC--BCS transition of Sm~$4f$-$5d$ exciton; 
therefore, a comprehensive understanding of these phenomena will help to elucidate the fruitful physical properties of SmS.

%%%%%%%%%%%%%%%%%%%%%%%%%%%%%%

%\section{Conclusion}
%
To summarize, we have investigated the electronic structure change of insulating \bSmS by the current application.
By analyzing the THz reflectivity spectra at temperatures below 100~K, the carrier density rapidly increased with current application up to 0.8~A and slowly after that, accompanied by the decreasing relaxation time, suggesting the changing of the carriers' character and the electronic structure at high currents.
The current application also increased the direct energy gap size at the $X$ point in the Brillouin zone and the increase of Sm ions' mean valence, suggesting that the applied current induces the Sm~$4f$-$5d$ hybridization.
The observed current-induced metallization accompanying the valence transition in \bSmS is a novel phenomenon of insulator-to-metal transitions.

%%%%%%%%%%%%%%%%%%%%%%%%%%%%%%
%\section*{Acknowledgments}
%
We appreciate Prof. T.~Ito for his fruitful discussion and UVSOR Synchrotron staff members for their support during synchrotron radiation experiments.
This work was partly performed under the Use-of-UVSOR Synchrotron Facility Program (Proposals Nos.~20-735, 21-677, 22IMS6015, 22IMS6029) 
of the Institute for Molecular Science, National Institutes of Natural Sciences.
This work was partly supported by JSPS KAKENHI (Grant No.~20H04453).

%%%%%%%%%%%%%%%%%%%%%%%%%%%%%%
%\begin{thebibliography}{99}
%
%
%\end{thebibliography}

%\bibliographystyle{plain}
\bibliographystyle{apsrev4-2}
%\bibliography{../../../bibtex/library}
\bibliography{SmS-I}

\end{document}

\title{
Supplementary Material for 
Current-Induced Metallization and Valence Transition in Black SmS%: \\Optical and Photoelectrical Studies
}
\author{Shin-ichi Kimura}
\email{kimura.shin-ichi.fbs@osaka-u.ac.jp}
\affiliation{Graduate School of Frontier Biosciences, Osaka University, Suita, Osaka 565-0871, Japan}
\affiliation{Department of Physics, Graduate School of Science, Osaka University, Toyonaka, Osaka 560-0043, Japan}
\affiliation{Institute for Molecular Science, Okazaki, Aichi 444-8585, Japan}
\author{Hiroshi Watanabe}
\affiliation{Graduate School of Frontier Biosciences, Osaka University, Suita, Osaka 565-0871, Japan}
\affiliation{Department of Physics, Graduate School of Science, Osaka University, Toyonaka, Osaka 560-0043, Japan}
\author{Shingo Tatsukawa}
\affiliation{Department of Physics, Graduate School of Science, Osaka University, Toyonaka, Osaka 560-0043, Japan}
\author{Takuto Nakamura}
\affiliation{Graduate School of Frontier Biosciences, Osaka University, Suita, Osaka 565-0871, Japan}
\affiliation{Department of Physics, Graduate School of Science, Osaka University, Toyonaka, Osaka 560-0043, Japan}
\author{Keiichiro Imura}
\thanks{Present address: Institute of Liberal Arts and Sciences, Nagoya University, Nagoya, Aichi 464-8601, Japan}
\affiliation{Department of Physics, Graduate School of Science, Nagoya University, Nagoya Aichi 464-8602, Japan}
\author{Hiroyuki S. Suzuki}
\affiliation{Institute for Solid State Physics, The University of Tokyo, Kashiwa, Chiba 277-8581, Japan}
\author{Noriaki K. Sato}
\thanks{Present address: Center for General Education, Aichi Institute of Technology, Toyota, Aichi 470-0392, Japan}
\affiliation{Department of Physics, Graduate School of Science, Nagoya University, Nagoya Aichi 464-8602, Japan}
\date{\today}
%
%
%\pacs{71.27.+a, 78.20.-e}% PACS, the Physics and Astronomy
%%%%%%%%%%%%%%%%%%%%%%%%%%%%%%%%%%%%%%%%%%%%%%%%%%%%%%%%%%%%
\maketitle
%
%%%%%%%%%%%%%%%%%%%%%%%%%%%%%%%%%%%%%%%%%%%%%%%%%%%%%%%%%%%%
\section*{S1.~Drude and Lorentz fitting for reflectivity spectra}
%
%%%%%%%%%%%%%% FIG. S1. Fitting results %%%%%%%%%%%%%%%%%%%%
\begin{figure}[t]
\begin{center}
\includegraphics[width=0.45\textwidth]{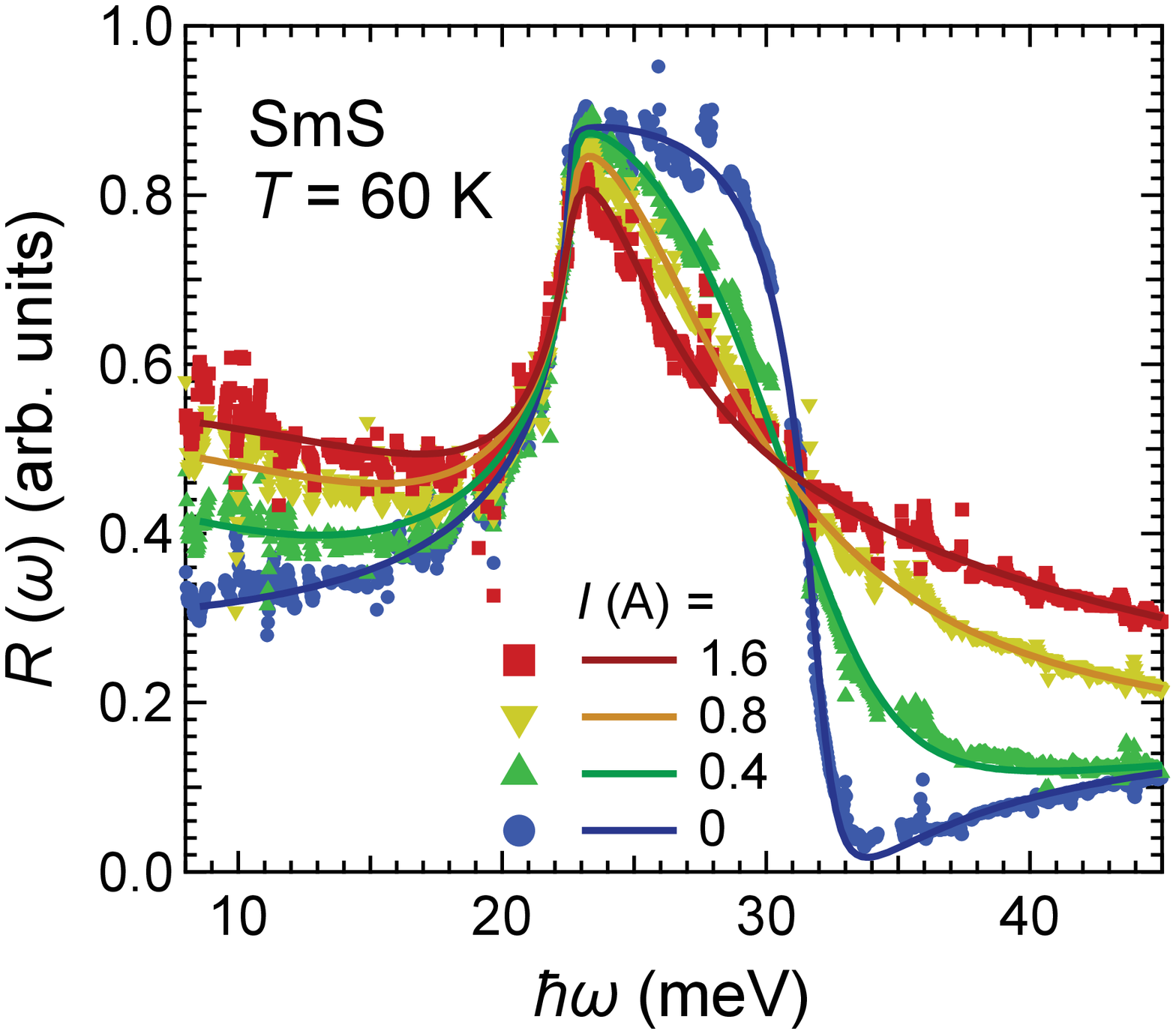}
\end{center}
\caption*{
%(Color online)
FIG. S1. Current-dependent reflectivity spectra (marks) at $T=60$~K and the Drude-Lorentz (DL) fitting results (solid lines).
The DL fitting was performed using a nonlinear least-square method expected as the combination of one Drude and one Lorentz component, as shown in Eq.~S1.
Here, $c$ was evaluated as 0.975.
}
\label{fig:phonon}
\end{figure}
%%%%%%%%%%%%%%%%%%%%%%%%%%%%%%%%%%%%%%%

In the current-dependent reflectivity (\R) measurements, the absolute value of the \R could not be obtained because the conventional {\it in-situ} gold evaporation method~\cite{Homes1993} can not be used.
Then, we adopted the following combination of Drude and Lorentz functions~\cite{Dressel2002} for the fitting of experimentally obtained \R spectra ($R_{expt}(\omega)$), which do not have exact reflectivity, as follows:
\begin{eqnarray*}
\tilde{\varepsilon}(\omega)=\varepsilon_\infty&-&\dfrac{e^2}{\varepsilon_0 m_0}\left(\dfrac{N_{eff}}{\omega^2+i\omega/\tau}
+\dfrac{N_{ph}}{(\omega^2-\omega_{ph}^2)+i\Gamma_{ph}\omega}\right) \nonumber \\
R_0(\omega)&=&\left|\dfrac{1-\tilde{\varepsilon}(\omega)^{1/2}}{1+\tilde{\varepsilon}(\omega)^{1/2}}\right|^2 \nonumber \\
R_{expt}(\omega)&=&cR_0(\omega) \hspace{30mm}{\rm (S1)}
\end{eqnarray*}
Here, $\tilde{\varepsilon}(\omega)$ is a complex dielectric function as a function of frequency (photon energy).
Constant values of $\varepsilon_\infty$, $\varepsilon_0$, $e$, and $m_0$ are the sum of the real part of the dielectric constant at a higher frequency (energy) region, the dielectric constant of vacuum, the elementary charge, and the electron rest mass, respectively.
In the Drude function, $N_{eff}$ is the effective electron density, $N_{eff}=\dfrac{N_0 m_0}{m^*}$, where $N_0$ and $m^*$ are the carrier density and the effective carrier mass, and $\tau$ is the relaxation time of carriers.
In the Lorentz function, $N_{ph}$ are the effective density of phonons, $N_{ph}=\dfrac{N_{ph}^0 m_0}{m_{ph}}$, where $N_{ph}^0$ and $m_{ph}$ are the density and reduced mass of the phonon, respectively, and $\omega_{ph}$ and $\Gamma_{ph}$ are the frequency and the dumping constant of the phonon, respectively.
$R_{expt}(\omega)$ spectra can be regarded as a relative \R spectra, which corresponds to actual \R spectra ($R_0(\omega)$) multiplied by the magnification constant ($c$).
We obtained seven parameters ($\varepsilon_\infty$, $N_{eff}$, $\tau$, $N_{ph}$, $\omega_{ph}$, $\Gamma_{ph}$, and $c$) in the function fitted by a nonlinear least-square method to $R_{expt}(\omega)$.
In these parameters, even though $c$ is changed in different sample settings, $\varepsilon_\infty$, and $\omega_{ph}$ are almost constant as 3.0 and 22.5~meV, respectively.
Other parameters, $N_{eff}$, $\tau$, $N_{ph}$, and $\Gamma_{ph}$, are changed by temperatures and currents.
An example of the fitting results is shown in Fig.~S1.
$N_{eff}$ and $\tau$ as functions of temperature and current are plotted in Figs.~1(c--f), and $N_{ph}$ and $\Gamma_{ph}$ are shown in the next section.

%%%%%%%%%%%%%%%%%%%%%%%%%%%%%%%%%%%%%%%%%%%%%%%%%%%%%%%%%%%%
\section*{S2.~Temperature and Current dependences of phonon parameters}

%%%%%%%%%%%%%% FIG. 2. Phono parameters %%%%%%%%%%%%%%%%%%%%
\begin{figure}[t]
\begin{center}
\includegraphics[width=0.45\textwidth]{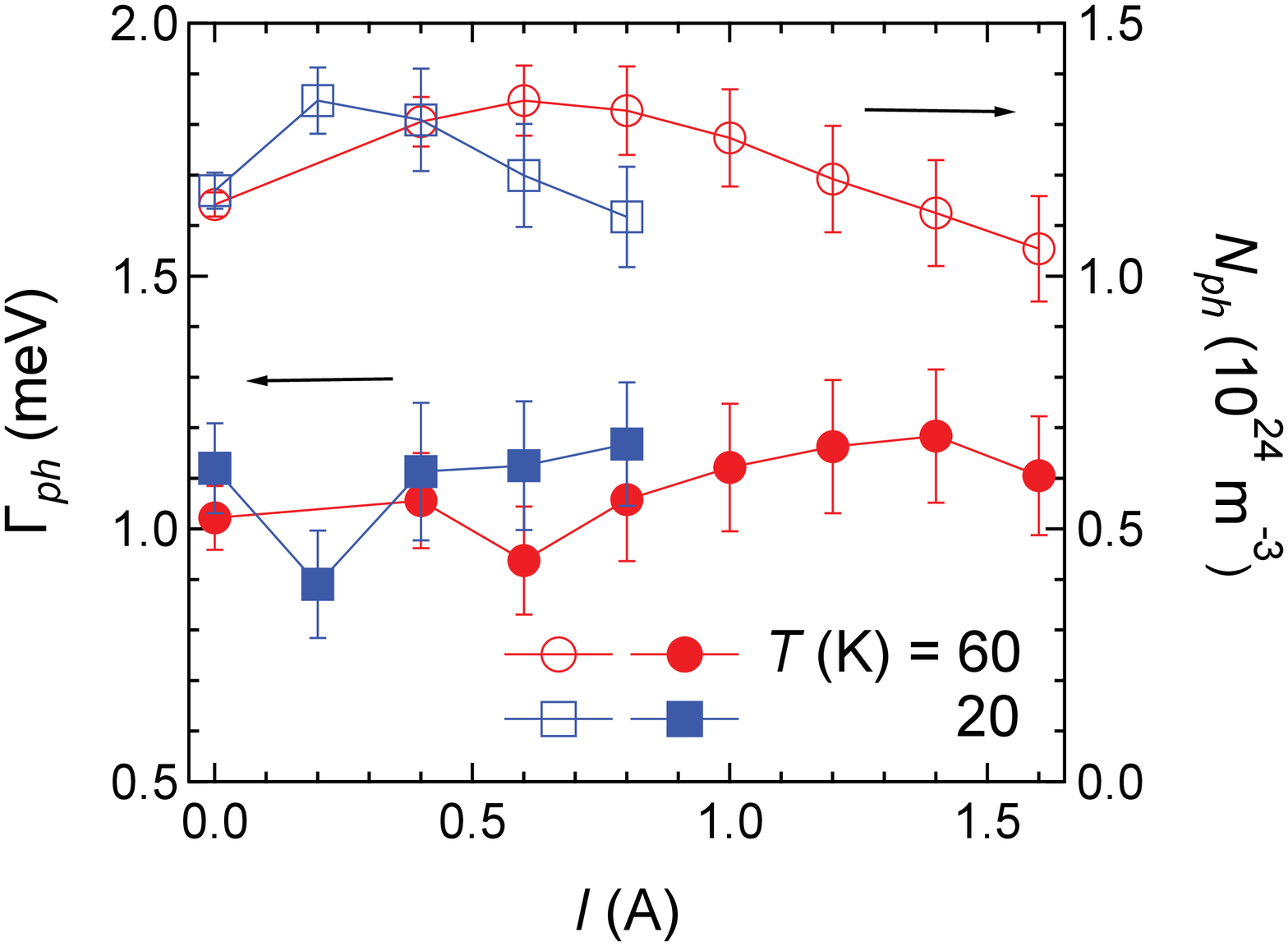}
\end{center}
\caption*{
%(Color online)
FIG. S2. Current-dependent effective phonon density $N_{ph}$ and dumping constant $\Gamma_{ph}$ of the TO phonon at about 20~meV at 20 and 60~K, at which temperature nonlinear current dependences of the carrier density and relaxation time are visible.
}
\label{fig:phonon}
\end{figure}
%%%%%%%%%%%%%%%%%%%%%%%%%%%%%%%%%%%%%%%

Current-dependent $\Gamma_{ph}$ and $N_{ph}$ for TO phonons at $T = 20$ and 60~K, at which the nonlinear current dependence appears, are shown in Fig.~S2.
Both parameters are unchanged so much in comparison with $N_{eff}$ and $\tau$ shown in Figs.~1(c--f), especially the damping constant $\Gamma_{ph}$ is almost constant in the whole current region, even though $N_{eff}$ and $\tau$ are strongly variable.
This result suggests that the sample temperature, one of the origins of the $\Gamma_{ph}$, did not change so much in the measured current region below 1.6~A.

%%%%%%%%%%%%%%%%%%%%%%%%%%%%%%

%\begin{thebibliography}{99}
%
%
%\end{thebibliography}

%\bibliographystyle{plain}
\bibliographystyle{apsrev4-2}
%\bibliography{../../../bibtex/library}
\bibliography{SmS-I}